\documentclass[epjST]{svjour}
\usepackage{graphics}
\usepackage{epsf,epsfig}
\def\xb{\overline{x}}

\def\vk{{\bf k}_{\perp}}

\begin{document}
\title{Transversity effects in light meson leptoproduction.}
\author{S.V.Goloskokov\fnmsep\thanks{\email{goloskkv@theor.jinr.ru}} }
\institute{Bogoliubov Laboratory of Theoretical  Physics,
  Joint Institute for Nuclear Research, Dubna, Russia}
\abstract{The light meson electroproduction is analyzed   within
the handbag approach. We consider the leading-twist contribution
together with the transversity  effects. We show that the
transversity Generalized Parton Distributions (GPDs), $H_T$ and
$\bar E_T$  are essential in the description of spin effects in
the pseudoscalar and vector meson leptoproduction.
} 
\maketitle
\section{Introduction}
\label{intro} The handbag approach where the amplitudes factorize
into the hard subprocess and  GPDs \cite{fact} was successfully
applied to the light vector meson  (VM) leptoproduction at high
photon virtualities $Q^2$ \cite{gk06} and the pseudoscalar meson
(PM) leptoproduction \cite{gk09}.

In the leading twist approximation the amplitudes of the PM
leptoproduction  are sensitive to the GPDs $\widetilde{H}$ and
$\widetilde{E}$.   It was found that these contributions  are not
sufficient to describe spin effects in the PM production at
sufficiently low $Q^2$ \cite{gk09}. The essential contributions
from the transversity GPDs  $H_T$, $\bar E_T$ are needed
\cite{gk11} to be consistent with experiment. Within the handbag
approach the transversity GPDs contribute to the amplitude
together with the twist-3 meson wave function. We discuss the
handbag approach and properties of meson production amplitudes in
 section 2.

In  section 3 of this report we study the experimental evidence of
transversity effects in the cross sections of the PM
leptoproduction at HERMES and CLAS energies. Our results are in
good agreement with experiment on the $\pi^0$ production. We show
that the transversity GPDs lead to a large transverse cross
section for most reactions of the pseudoscalar meson production
\cite{gk11} which has the twist-3 nature.

We investigate the role of transversity GPDs in the VM
leptoproduction in  section 3  within the handbag approach
\cite{gk13}. The importance of the transversity GPDs was found in
the SDMEs and in asymmetries measured with a transversely
polarized target. The obtained results are in good agreement with
CLAS, HERMES and COMPASS data.

\section{Handbag approach. Properties of meson production amplitudes}
Within the handbag approach the meson production amplitude is
factorized  at sufficiently high $Q^2$ \cite{fact} into a hard
subprocess amplitude ${\cal H}$ and  GPDs $F$ which contain
information on the hadron structure see, Fig.1.

The subprocess amplitude is calculated within the MPA
\cite{sterman}. We consider the power $k_\perp^2/Q^2$ corrections
in the propagators of the hard   subprocess  ${\cal H}$ together
with the nonperturbative $\vk$-dependent meson wave function
\cite{koerner}. The power corrections can be regarded as an
effective consideration of the higher twist effects. The gluonic
corrections are treated in the form of the Sudakov factors whose
resummation can be done in the impact parameter space
\cite{sterman}.
\begin{figure}
\centering \mbox{\epsfysize=30mm\epsffile{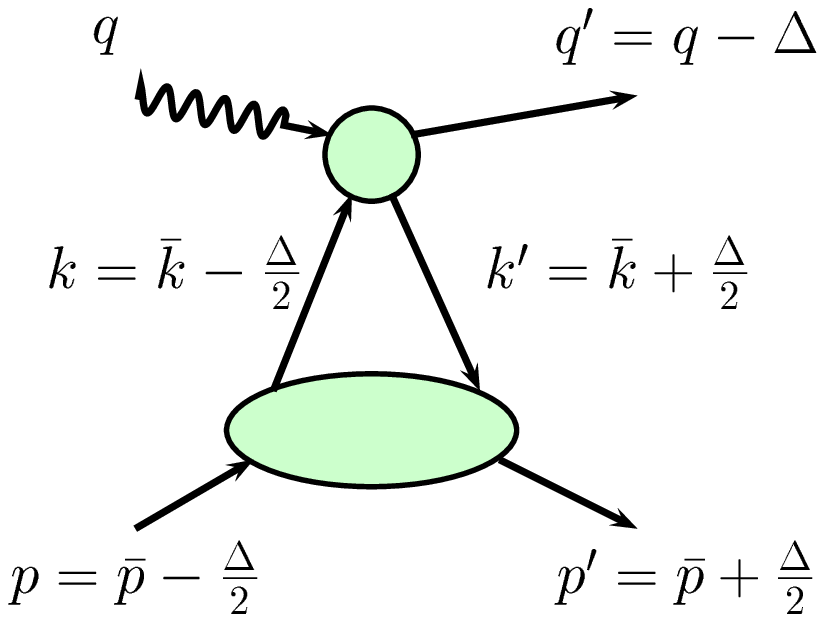}}
\caption{The handbag diagram for the meson electroproduction off
proton.} \label{kt_h}
\end{figure}

The leading contributions to the proton non-flip amplitude can be
expressed in terms various parton effects
\begin{equation}\label{ff}
 {\cal M}_{\mu' +,\mu +} \propto \int_{-1}^1 dx
   {\cal H}^a_{\mu' +,\mu +} F^a(x,\xi,t).
\end{equation}
Here $a$ is a flavor factor. In the VM production we have $F^a$
GPDs contributions  from gluons, quarks and sea. In PM production
polarized GPDs $\tilde F^a$ contributes.

The GPDs are estimated using the double distribution
representation \cite{mus99} which connects  GPDs $F$ with PDFs $h$
through the double distribution function $\omega$. For the valence
quark contribution it looks like
\begin{equation}\label{ddf}
\omega_i(x,y,t)= h_i(x,t)\,
                   \frac{3}{4}\,
                   \frac{[(1-|x|)^2-y^2]}
                           {(1-|x|)^{3}}.
\end{equation}
 The  $t$- dependence in PDFs $h$ is considered in the Regge form
\begin{equation}\label{pdfpar}
h(\beta,t)= N\,e^{b_0 t}\beta^{-\alpha(t)}\,(1-\beta)^{n},
\end{equation}
and $\alpha(t)$ is the corresponding Regge trajectory.  The
parameters in (\ref{pdfpar}) are obtained from the known
information about PDFs \cite{CTEQ6} e.g, or from the nucleon form
factor analysis \cite{pauli}.

At the leading-twist accuracy the PM production  is only sensitive
to the GPDs $\widetilde{H}$ and $\widetilde{E}$ which contribute
to the amplitudes for longitudinally polarized virtual photons
\cite{gk09}. It was found that at low $Q^2$ data on the PM
leptoproduction also require the contributions from the
transversity GPDs  $H_T$ and $\bar E_T$ which determine the
amplitudes $M_{0-,++}$ and $M_{0+,++}$ respectively. Within the
handbag approach the transversity GPDs are accompanied by a
twist-3 meson wave function in the hard amplitude ${\cal H}$
\cite{gk11} which is the same for  both the ${\cal
M}^{M,tw-3}_{0\pm,++}$ amplitudes in (\ref{ht}).
\begin{equation}\label{ht}
{\cal M}^{M,tw-3}_{0-,++} \propto \,
                            \int_{-1}^1 d\xb
   {\cal H}_{0-,++}(\xb,...)\,H^{M}_T;\;
   {\cal M}^{M,tw-3}_{0+,++} \propto \, \frac{\sqrt{-t'}}{4 m}\,
                            \int_{-1}^1 d\xb
 {\cal H}_{0-,++}(\xb,...)\; \bar E^{M}_T.
\end{equation}

The $H_T$ GPDs are connected with transversity PDFs  as
\begin{equation}
  H^a_T(x,0,0)= \delta^a(x);\;\;\; \mbox{and}\;\;\;
\delta^a(x)=C\,N^a_T\, x^{1/2}\, (1-x)\,[q_a(x)+\Delta q_a(x)].
\end{equation}
We parameterize the PDF $\delta$ by using the model \cite{ans}.

The information on $\bar E_T$ is obtained now only in the lattice
QCD \cite{lat}. The lower moments of $\bar E_T^u$ and $\bar E_T^d$
were found to be quite large, have the same sign and a similar
size. As a result,  we have large $\bar E_T$ contributions to the
$\pi^0$ production.

For the VM production the transversity $M_{0-,++}$ and $M_{0+,++}$
amplitudes have the form (\ref{ht}) but they are parametrically
about 3 times smaller \cite{gk13} with respect to the PM case. In
calculations  we use the same parameterizations for transversity
GPDs $H_T$ and $\bar E_T$ which were obtained in our study of the
PM leptoproduction and can be found in \cite{gk11,gk13}. The
double distribution  is used to calculate GPDs in all cases.

\section{Transversity effects in  meson leptoproduction}
In this section, we present our results on the PM and the VM
leptoproduction based on the handbag approach. In calculation, we
use the leading contribution (1) together with the transversity
effects (\ref{ht}) which are essential at low $Q^2$.
 In Fig. 2 (left), we present the model results for the $\pi^0$
 production cross section.
The transverse cross section, where the $\bar E_T$ and $H_T$
contributions  are important \cite{gk11}, dominates at  low $Q^2$.
At small momentum transfer the $H_T$ effects are visible and
provide a nonzero cross section. At $-t' \sim 0.2 \mbox{GeV}^2$
the $\bar E_T$ contribution becomes essential in $\sigma_T$ and
gives a maximum in the cross section. A similar contribution from
 $\bar E_T$ is observed in the interference cross section
$\sigma_{TT}$ \cite{gk11}. The fact that we describe well both
unseparated $\sigma=\sigma_{T}+\epsilon \sigma_{L}$ and
$\sigma_{TT}$ cross sections can probably indicate that the
transversity effects were observed at CLAS \cite{bedl}.

\begin{figure}[h!]
\begin{center}
\begin{tabular}{cc}
\includegraphics[width=6.3cm,height=5.1cm]{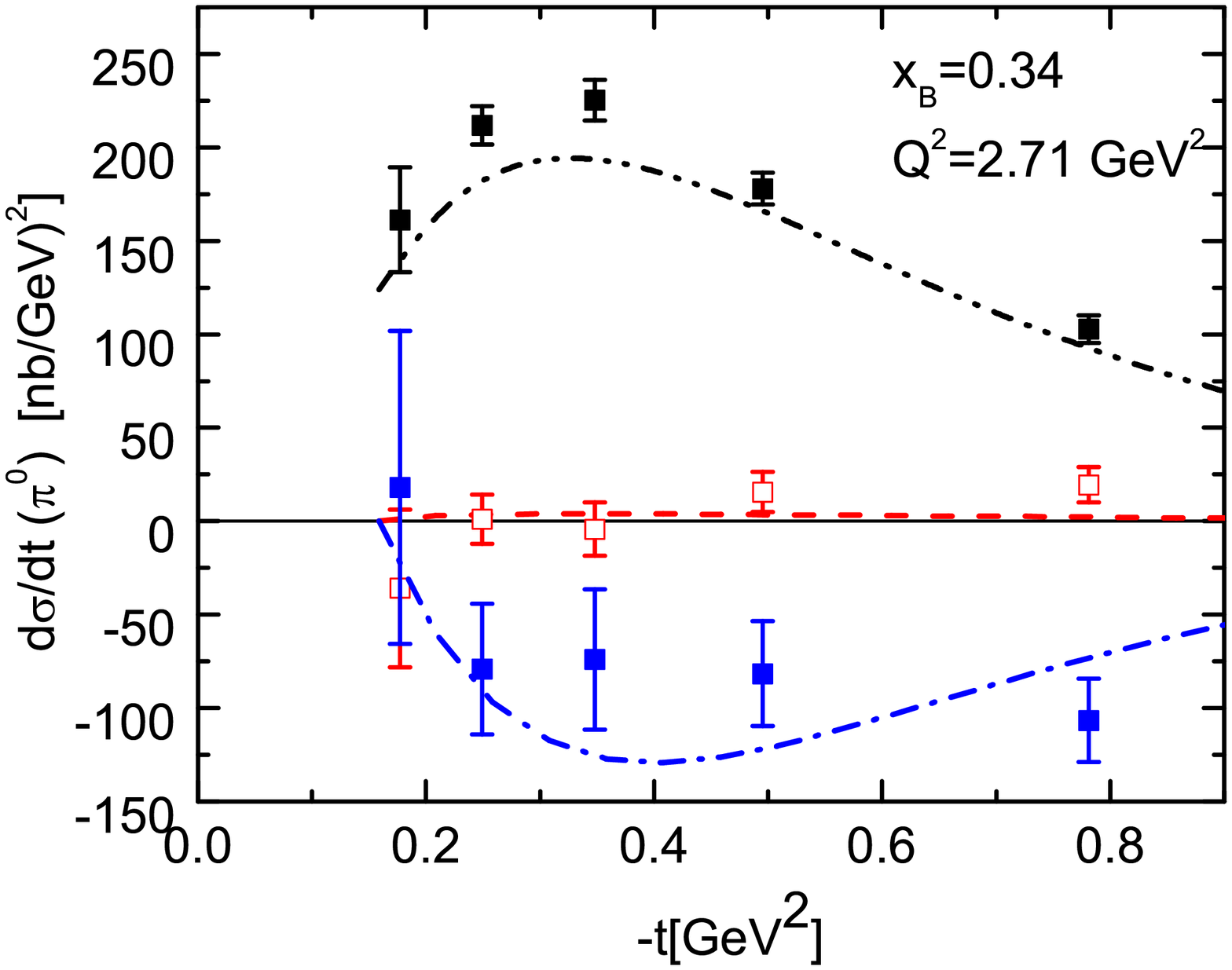}&
\includegraphics[width=6.cm,height=5.1cm]{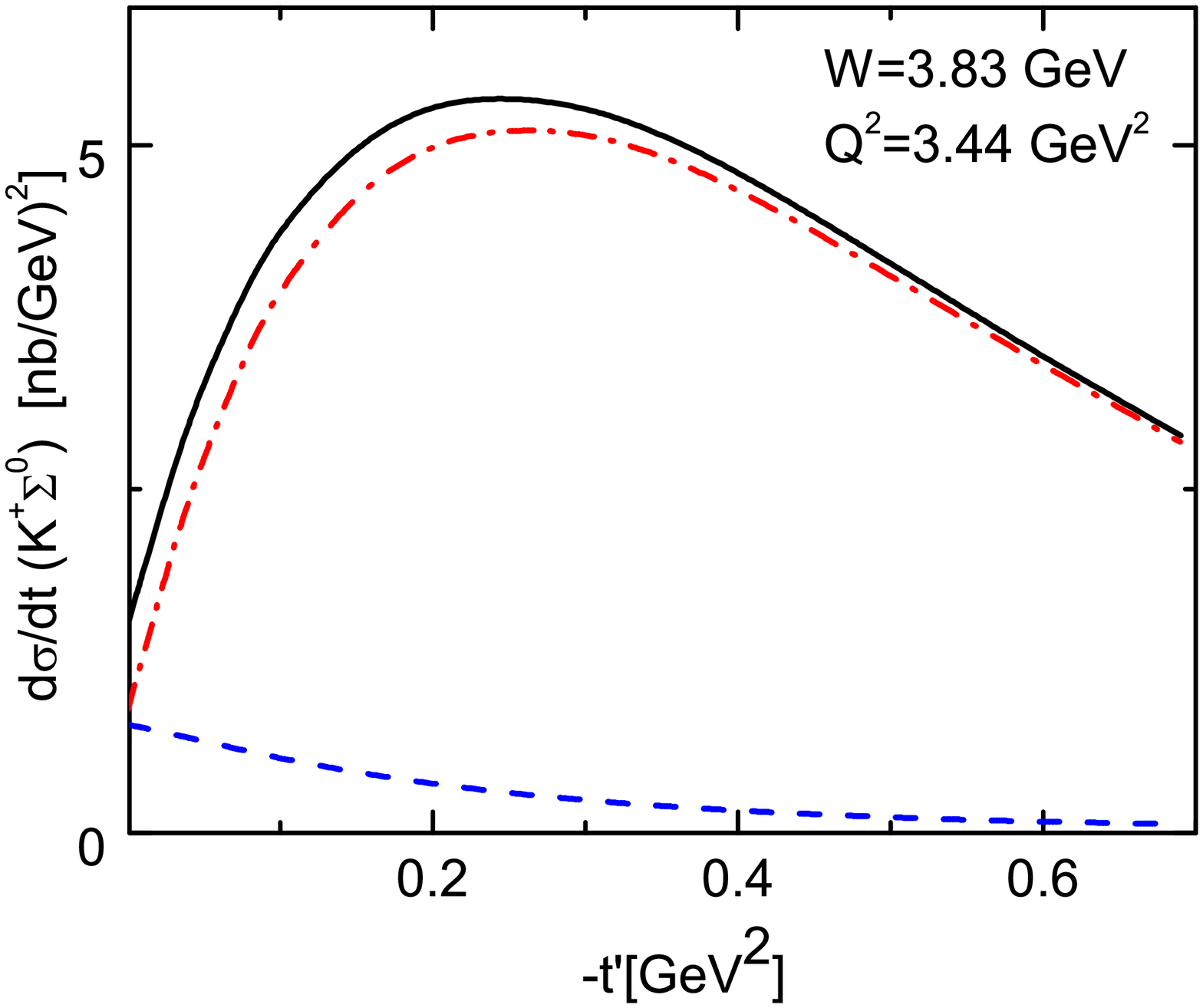}
\end{tabular}
\end{center}
\caption{Left: $\pi^0$ production in the CLAS energy range
together with the data \cite{bedl}. Dashed-dot-dotted line-
$\sigma=\sigma_{T}+\epsilon \sigma_{L}$, dashed
line-$\sigma_{LT}$, dashed-dotted- $\sigma_{TT}$. Right:
 Cross sections of the $K^+ \Sigma^0$ production
at HERMES energies.}
\end{figure}

Using the same model approach the kaons leptoproduction was
considered. We show the model results  \cite{gk11} for the $K^+
\Sigma^0$ leptoproduction in Fig. 2 (right). Both $H_T$ and $\bar
E_T$ transversity GPDs contribute here and provide a cross section
similar  in  form  to the $\pi^0$ case. In both reactions the
leading-twist longitudinal cross section is small, see Fig. 2
(right), at low $Q^2$ where the transverse cross section
dominates. At sufficiently high $Q^2$ the $\sigma_{T}$ cross
section which decreases rapidly with $Q^2$ will be small with
respect to the leading twist  $\sigma_{L}$.

\begin{figure}[h!]
\begin{center}
\begin{tabular}{cc}
\includegraphics[width=6.1cm,height=5.1cm]{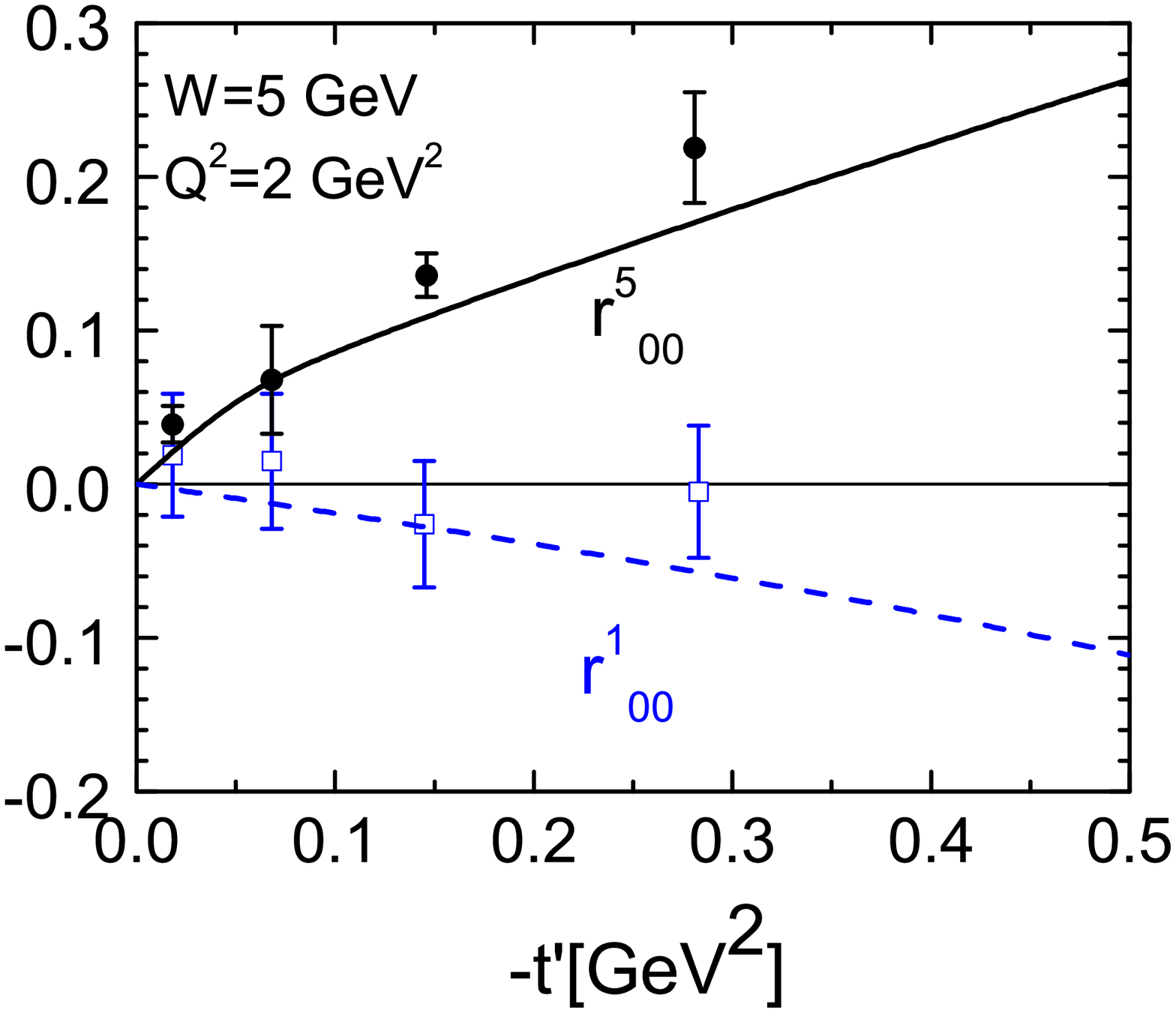}&
\includegraphics[width=6.1cm,height=5.1cm]{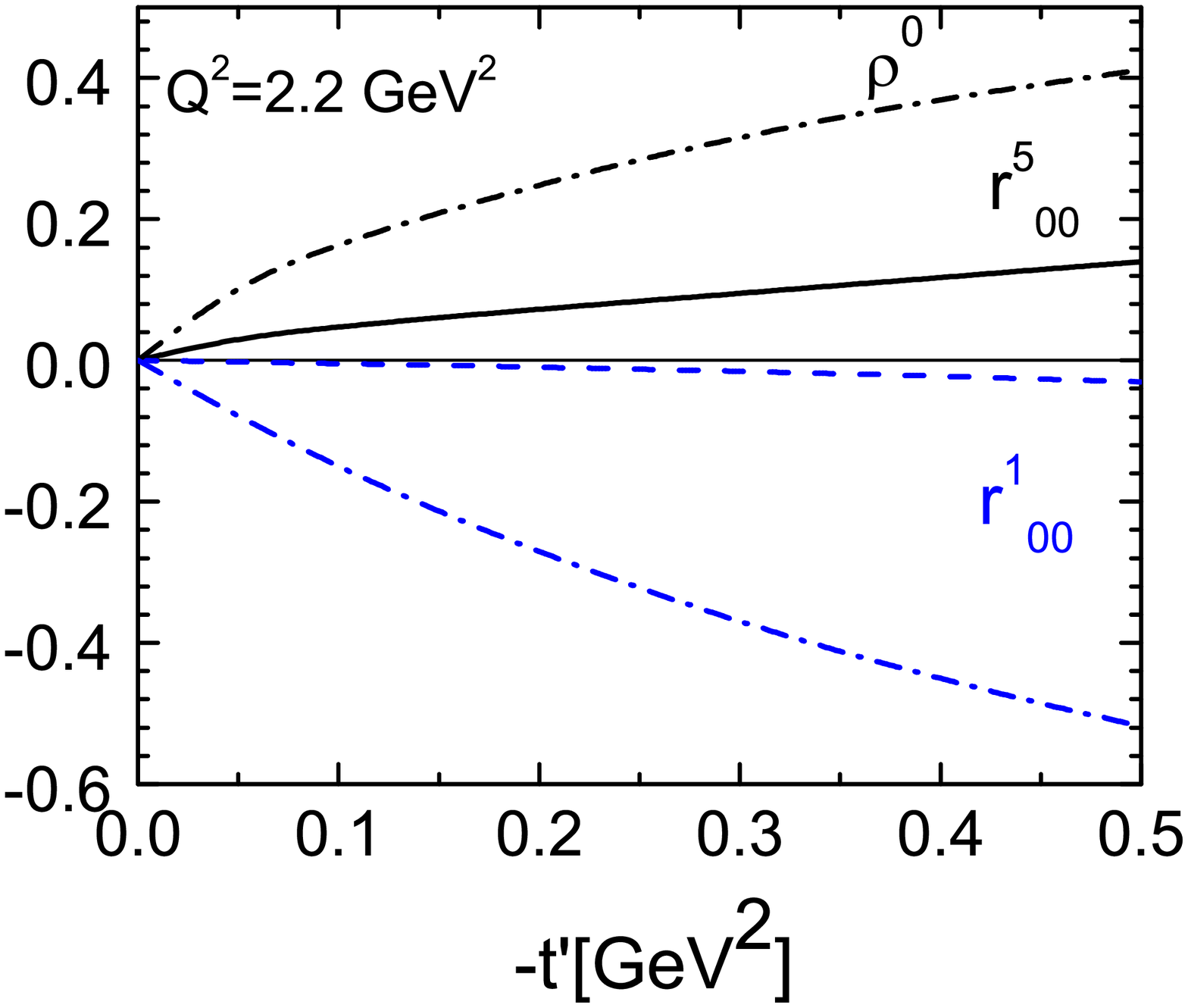}
\end{tabular}
\end{center}
\caption{Transversity effects in the $r^5_{00}$, $r^1_{00}$ SDMEs.
Left: at $W=5 \mbox{GeV}$ together with HERMES data \cite{airap}.
Right: SDMEs at CLAS (dot-dashed line) and COMPASS energies}
\end{figure}

Now we shall consider the effects of transversity in the  VM
production. We use here the same GPDs as for the PM case. The
importance of the transversity GPDs was examined in  the SDMEs and
in the transversely polarized target asymmetries. The
$M_{0+,++}=<\bar E_T>$ amplitude is essential in some SDMEs.
Really,
\begin{equation}\label{sdme}
r^5_{00} \sim \mbox{Re}[M_{0+,0+}^* M_{0+,++}];\;\;\; r^1_{00}
\sim -|M_{0+,++}|^2;\;\;\;r^{04}_{10} \sim \mbox{Re}[M_{++,++}^*
M_{0+,++}].
\end{equation}

Our results   for these the $r^5_{00}$, $r^1_{00}$ SDMEs  in the
$\rho^0$ meson production at HERMES are shown in  Fig. 3(left).
The values and signs are in good agreement with HERMES
experimental data \cite{airap}. The $r^{04}_{10}$ SDME is
reproduced well too. We observe that large $\bar E_T$ effects
found in the $\pi^0$ channel are compatible with SDME of the
$\rho$ production at HERMES energies. The model prediction
\cite{gk13} for these SDMEs at CLAS and COMPASS energies are shown
in Fig 3 (right). The predicted SDMEs are quite large and can be
measured experimentally.

\begin{figure}[h!]
\begin{center}
\begin{tabular}{cc}
\includegraphics[width=6.1cm,height=5.1cm]{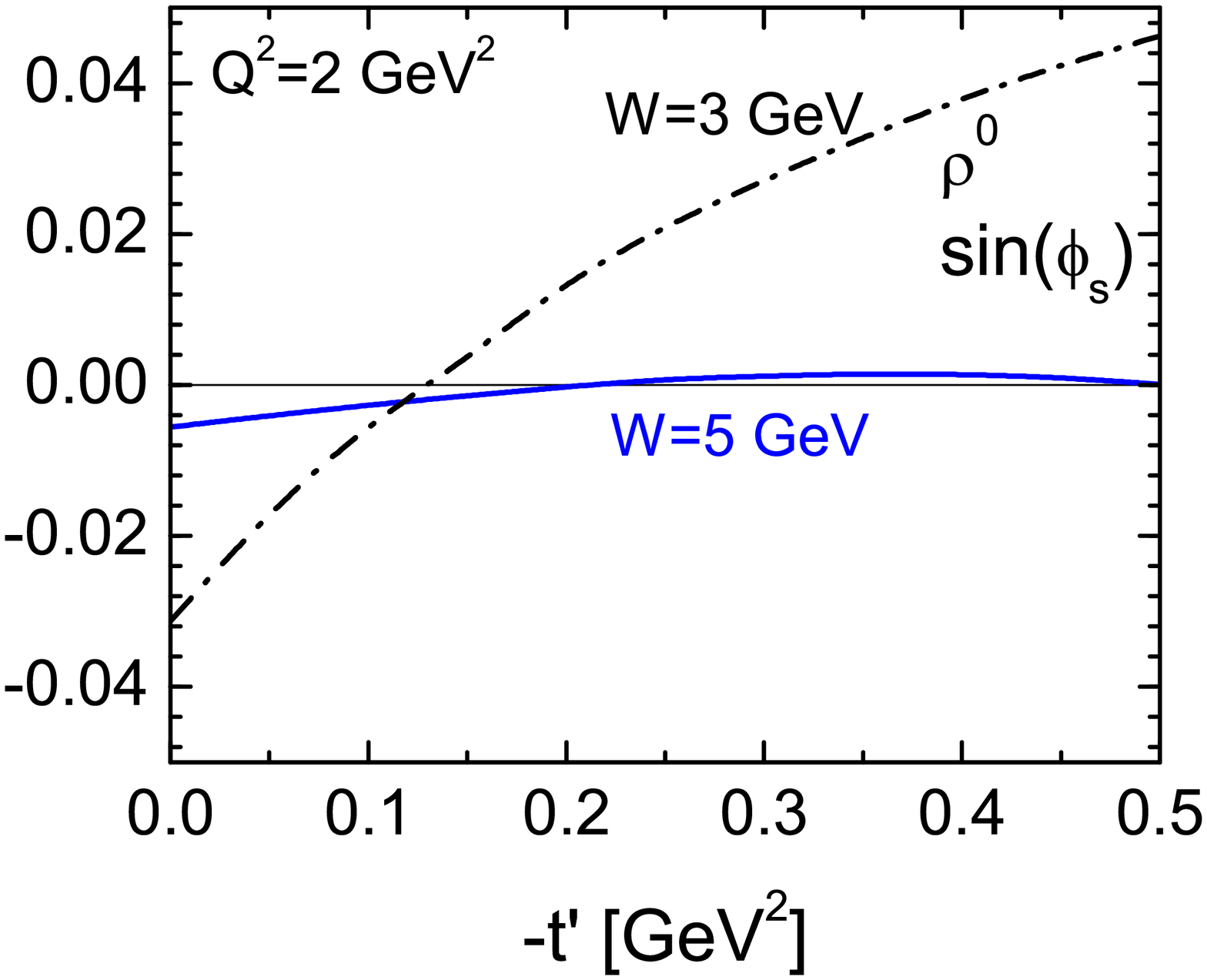}&
\includegraphics[width=6.1cm,height=5.3cm]{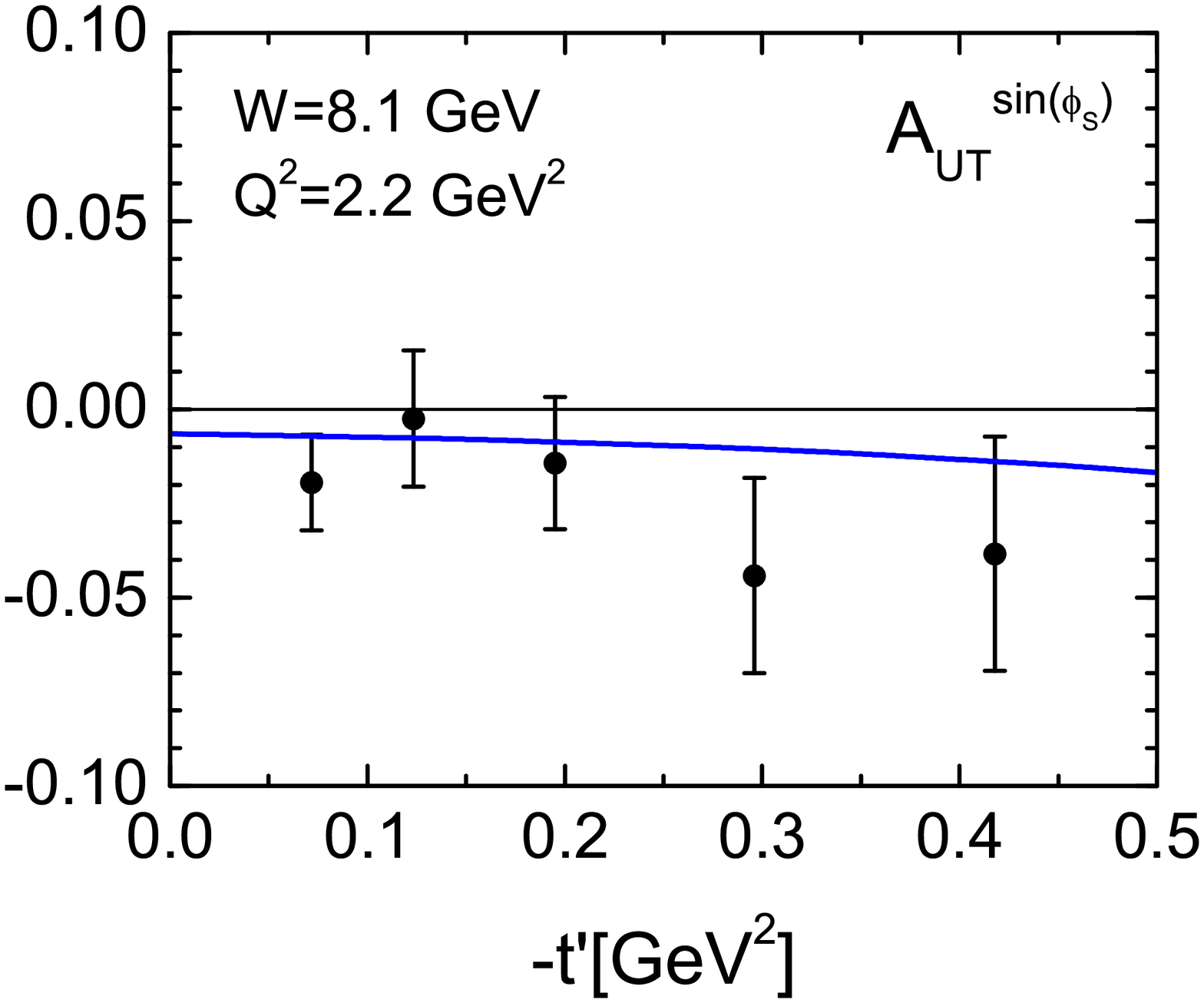}
\end{tabular}
\end{center}
\caption{Model results for the $A_{UT}^{\sin(\phi_s)}$ asymmetry.
 Left: at CLAS and HERMES energies. Right: at COMPASS energies. Data are from
\cite{autcomp13}.}
\end{figure}

The transversity effects in $A_{UT}$, $A_{LT}$ asymmetries was
analyzed in \cite{gk13} too.  The $\sin (\phi_s)$ moment of the
$A_{UT}$ asymmetry is determined by the interference transversity
$H_T$ contribution and the non-flip amplitude $M_{0+,0+}$.

\begin{equation}\label{sinfs}
 A_{UT}^{\sin(\phi_s)} \sim \mbox{Im}[M_{0-,++}^*
M_{0+,0+}]; \;\;\;M_{0-,++}=<H_T>
\end{equation}
 It was found that the asymmetry is
not small. The energy dependence of the $A_{UT}^{\sin(\phi_s)}$
asymmetry from CLAS to HERMES is quite rapid and shown in Fig. 4
(left). This prediction can be verified at a future CLAS
experiment to test the $x$- dependence of GPDs $H_T$. At COMPASS
energies our results are shown in  Fig. 4 (right). In  Fig. 5
(left) the $Q^2$ dependence of $A_{UT}^{\sin(\phi_s)}$ is
presented. Our results at COMPASS are compatible with the data
\cite{autcomp13}.

\begin{figure}[h!]
\begin{center}
\begin{tabular}{cc}
\includegraphics[width=6.1cm,height=5cm]{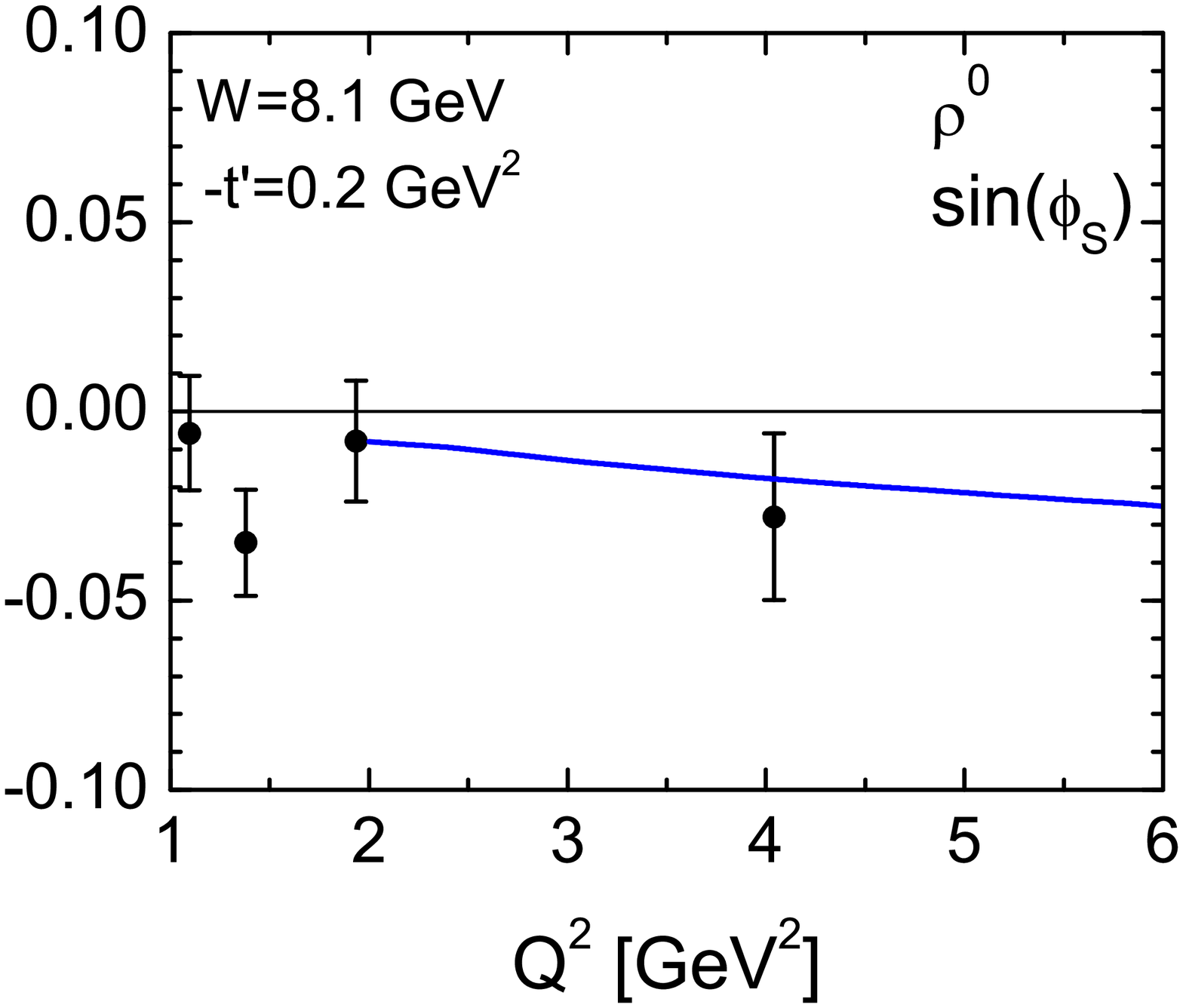}&
\includegraphics[width=6.1cm,height=5cm]{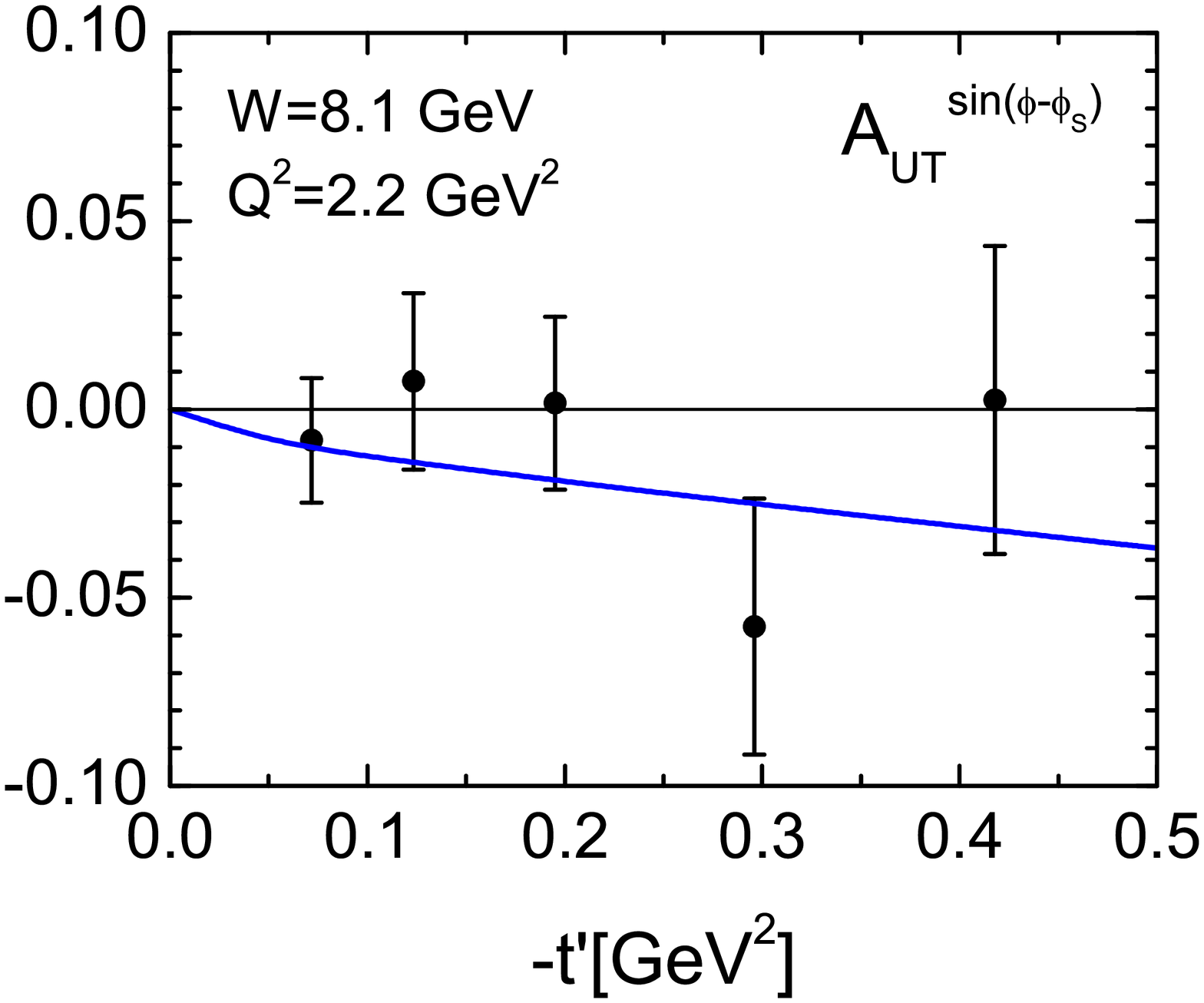}
\end{tabular}
\label{fig:4}
\end{center}
\caption{$Q^2$ dependencies of Left: $A_{UT}^{\sin(\phi_s)}$
asymmetry.  Right: $A_{LT}^{\cos(\phi_s)}$ asymmetry at COMPASS
together with data \cite{autcomp13}}
\end{figure}

In Fig. 5 (right) we show  our results for the $\sin(\phi-\phi_s)$
moment of the $A_{UT}$ asymmetry
\begin{equation}\label{autfmfs}
  A_{UT}^{\sin(\phi-\phi_s)} \sim \mbox{Im}[M_{0-,0+}^* M_{0+,0+} -
M_{0-,++}^* M_{0+,++}]
\end{equation}
at  COMPASS energies which are compatible with experiment. This
asymmetry is determined essentially by the interference of the
$<E>$ and $<F>$ contributions (\ref{ff}) and is consistent with
the data. The effects of transversity  is quite small here. We
describe well \cite{gk13} the $A_{UT}^{\sin(\phi-\phi_s)}$
asymmetry at HERMES energies \cite{rostom} too.

\section{Conclusion}
The exclusive electroproduction of vector and pseudoscalar mesons
was analyzed here within the  handbag approach where the amplitude
factorized in two parts. The first one is  the subprocess
amplitudes which are calculated using the $k_\perp$ factorization
\cite{sterman}. The other essential ingredients are the GPDs which
contain information about the hadron structure. The results based
on this approach on the cross sections and various spin
observables were found to be in good agreement with data at
HERMES, COMPASS and HERA energies at high $Q^2$ \cite{gk06}.

The leading-twist accuracy is not sufficient to describe the PM
leptoproduction at rather low $Q^2$. It was also found   that
rather strong twist 3 contributions  are required by experiment.
In the handbag approach they are determined by  the transversity
GPDs $H_T$ and $\bar E_T$ in convolution with a twist-3 pion wave
function. It turned out that the transversity GPDs lead to a large
transverse cross section for most reactions of the PM production.
There are some indications that the large transversity effects are
available now at CLASS \cite{bedl}.

In the VM leptoproduction the transversity GPDs  occur in the
amplitudes with a transversely polarized virtual photon and a
longitudinal polarized vector meson. The structure of the twist-3
amplitudes in the VM production is similar to the PM case. We show
the importance of the transversity GPDs in the SDMEs $r^5_{00}$,
$r^1_{00}$ and $r^{04}_{10}$ of $\rho^0$ production which were
found to be essentially dependent  on the $\tilde E_T$
transversity affects. Our results are in good agreement with
HERMES experimental data \cite{airap}.

The asymmetries measured with a transversely polarized target were
estimated as well.  The $A_{UT}^{\sin(\phi-\phi_s)}$ transverse
target spin asymmetry \cite{gk13}  is consistent with HERMES and
COMPASS data \cite{rostom,autcomp13}. We find a large $H_T$
contribution to the $A_{UT}^{\sin (\phi_s)}$ asymmetry which is
not small  at COMPASS \cite{gk13} and also compatible with the
experimental data \cite{autcomp13}. Our predictions were compared
by COMPASS \cite{autcomp13} with the COMPASS experimental data.
Thus, we can conclude that the transversity GPDs are extremely
essential in understanding the spin effects in the light meson
leptoproduction at moderate $Q^2$.
\bigskip

This work is supported  in part by the Russian Foundation for
Basic Research, Grant  12-02-00613  and by the Heisenberg-Landau
program.

\end{document}